# Optimisation of the design for the LOFT Large Area Detector Module

D. Walton[*a], B. Winter[a], S. Zane[a], T. Kennedy[a], A.J. Coker[a], M. Feroci[b,c], J.-W. Den Herder[d], A. Argan[e], P. Azzarello[h], D. Barret[f,g], L. Bradley[a], F. Cadoux[h], A. Cros[f], Y. Evangelista[b], Y. Favre[h], G. Fraser[i**], M.R. Hailey[a], T. Hunt[a], A. Martindale[i], F. Muleri[b], L. Pacciani[b], M. Pohl[h], P. Smith[a], A. Santangelo[j], S. Suchy[j], C. Tenzer[j], G. Zampa[k], N. Zampa[k]

on Behalf of the LOFT Consortium

[a]Mullard Space Science Laboratory, UCL, Holmbury St Mary, Dorking, Surrey, RH56NT,UK; [b]INAF-IAPS-Roma via Fosso del Cavaliere, 100, 00133, Rome, Italy; [c]INFN, Sez. Roma Tor Vergata, Via della Ricerca Scientifica 1 - 00133 Rome, Italy; [d]SRON, The Netherlands Institute of Space Research, Utrecht, The Netherlands; [e]INAF HQ, Viale del Parco Mellini 84, 00136, Rome, Italy; [f]CNES, IRAP, 9 Avenue du Colonel Roche, BP44346, 31028, Toulouse, France; [g]Universite de Toulouse; UPS-OMP; IRAP; Toulouse, France; [h]DPNC, Geneva University, Quai Ernest-Ansermet 24, CH-1211,Geneva, Switzerland, [i]Space Research Centre, Department of Physics and Astronomy, University of Leicester, Leicester, LE17RH, UK, [j]IAAT, University of Tuebingen, Sand 1, 72076, Tuebingen, Germany, [k]Istituto Nazionale di Fisica Nucleare, INFN, Sezione di Trieste, Padriciano 99, I-34149, Trieste, Italy.

## ABSTRACT

LOFT (Large Observatory for X-ray Timing) is an X-ray timing observatory that, with four other candidates, was considered by ESA as an M3 mission (with launch in 2022-2024) and has been studied during an extensive assessment phase. Its pointed instrument is the Large Area Detector (LAD), a 10 $m^2$-class instrument operating in the 2-30 keV range, which is designed to perform X-ray timing of compact objects with unprecedented resolution down to millisecond time scales.

Although LOFT was not downselected for launch, during the assessment most of the trade-offs have been closed, leading to a robust and well documented design that will be reproposed in future ESA calls. The building block of the LAD instrument is the Module, and in this paper we summarize the rationale for the module concept, the characteristics of the module and the trade-offs/optimisations which have led to the current design.

## 1. INTRODUCTION

High time resolution X-ray observation of compact objects is a powerful tool to probe particle physics not testable in terrestrial laboratories, such as matter and radiation behaviour in strong field gravity and at supra-nuclear density.

LOFT, the Large Observatory for X-ray Timing [1], was one of the candidates originally considered by ESA as an M3 mission, with the Large Area Detector (LAD) instrument providing the required combination of very large collecting area, accurate high-resolution timing information and fine spectral resolution [2]. The LOFT payload consists of two instruments (Figure 1): the LAD, a collimated spectrometer with a 10 $m^2$ effective area (on the 6 deployable panels in the baseline consortium configuration), and the Wide Field Monitor [3], a coded-mask detector that monitors a large fraction of the sky for changes in the state of targets of interest (NSs, BHs, and AGNs) or discovery and localization of new sources. The ground-breaking characteristic of the LAD is the large collecting area, which is 20 times larger than that of the best past timing missions (such as RXTE, the largest predecessor), allowing to study the X-ray variability of compact objects with an unprecedented resolution. The LAD is designed to operate in the energy range 2-30 keV (up to 80 keV in expanded mode) with good spectral resolution (<260 eV @ 6 keV Full Width Half Maximum, FWHM) and a temporal resolution of 10μs.

[*]d.walton@ucl.ac.uk; phone 0044 1483204190; fax 0044 1483 278312.

[**] This work is dedicated to the memory of Prof. George Fraser, a dear friend and esteemed colleague.

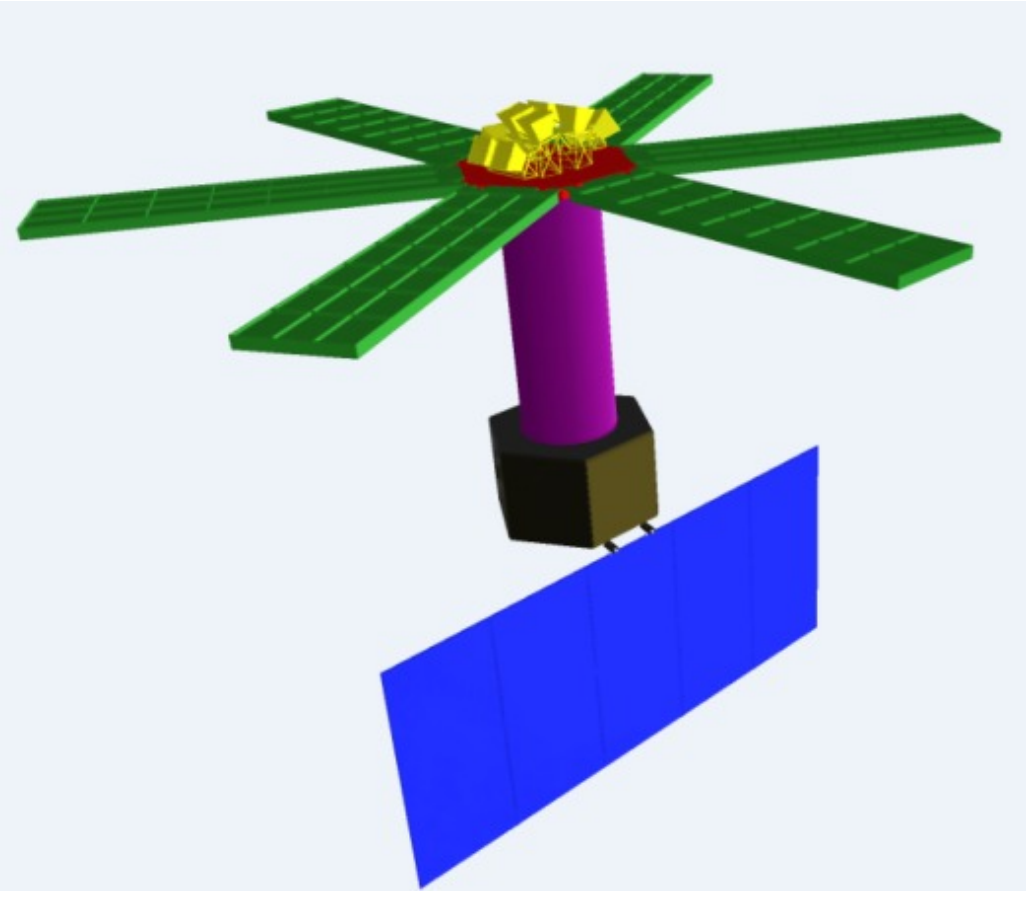

Figure 1 The configuration for LOFT as in the M3 proposal. Green = LAD, yellow = WFM, Red = Optical bench, Purple = Structural Tower, Gold = Bus, Blue = Solar array.

LAD achieves its 20x increase in effective area within a practicable design through two key technologies:

- large-area Silicon Drift Detectors (SDDs; [4]) designed on the heritage of the ALICE experiment at CERN/LHC;
- MicroPore Optic (MPO) collimators (one per SDD) based on lead-glass micro-capillary plates (the mechanical structure of the well-known microchannel plates).

The effective area requirement leads to a geometric area of ~18 $m^2$ or an active area of ~15 $m^2$.

The current baseline for the configuration, assumed by the consortium, is based on 6 detector panels (approximately 1x3 $m^2$ each), connected by hinges to an optical bench at the top of a tower (see Figure 1 and Figure 2). The basic LAD detection element is composed of SDD + Front-End Electronics (FEE) + Collimator. The 6 Detector Panels (DP) will be tiled with 2016 detectors, electrically and mechanically organized in groups of 16, referred to as Modules (Figure 3 and Figure 4). The read-out electronics are organized as follows: the FEEs of the 16 Detectors in a Module converge into a single Module Back End Electronics (MBEE, made up of two sections). One Panel Back-End Electronics (PBEE) for each DP is in charge of interfacing in parallel the 21 MBEEs on the DP, making the Module the basic redundant unit. For this reason, a careful optimization of the design of the Module (the building block of the LAD) is of utmost importance in order to reach the required performance in the weight and size available.

In this paper we will focus on the Module design and describe the main trade-offs that lead to the currently proposed configuration. The overall instrument design is described in more detail in [1] and [2].

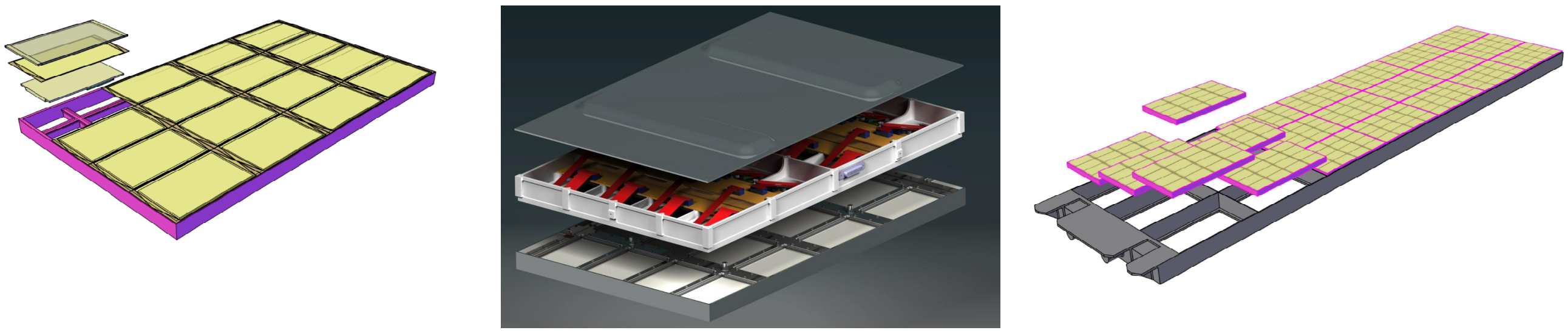

Figure 2 Left: Front-side view of a Module, showing the mounting of the collimator, SDD and the Front End Electronics. Center: Back-side view of a Module, showing the Module Back End Electronics. Right: a LOFT Detector Panel with all the assembled Modules and the interfaces to the deployment system.

## 2. MODULE RATIONALE, SYSTEM-LEVEL CONSIDERATIONS

In order to achieve an instrument of the scale of LAD, detailed consideration has to be given to optimizing all aspects of the implementation, e.g. minimising complex or time-consuming tasks which must be repeated on many units. This approach has to be applied to

- Manufacture, assembly and integration
- Testing, on the bench and in vacuum (including calibration)
- Transport (LAD requires significant amounts of transport of units across Europe)
- Assembly and (if required) replacement of sub-sections at higher levels of integration

These factors lead to an approach where the instrument is designed as a set of identical modules. An additional benefit of the modular approach is scaleability, the ability to achieve changes in the overall collecting area relatively simply by changing the number of modules.

The hierarchy/modularity of the LAD design is driven primarily by the following factors:

1. At the largest scale, instrument level, the scientific requirement for an effective area of ~10 sq. m at 8 keV, resulting in a geometric area of ~18 sq. m (to allow for pore wall absorption, collimator-to-collimator and module-to-module butting loss and alignment effects).
2. At the smallest scale, detector level, the practical size achievable of SDDs and collimator tiles (e.g. see [4] for information on the LOFT SDDs).
3. At upper- intermediate scale, the division of the instrument collecting area into individual panels, driven by spacecraft-level issues such as accommodation on the launcher, deployability, intra- and inter-panel alignment (as mentioned above, the consortium design has six panels, chosen to allow accommodation in a fairing).
4. At lower-intermediate scale, the trade-offs in practicalities of manufacture, power distribution, signal flow and bottlenecks, alignment, testability and replaceability: this is the design trade-off space where the optimization of the module design is achieved. As described above, the consortium design has 126 modules, with 16 detectors per module.

In the following sections, the trade-offs which have led to the current module design will be described.

In principle, LAD's ~2000 detectors (2016 in the consortium design) could be mounted individually in the panels. This might save some structural mass compared with the current design, but would severely compromise manufacturability, AIV, testability, calibration and the replaceability of defective parts.

Conceptually, creating an M-level hierarchy for N objects, a 'natural' division of complexity occurs when a unit at each level branches to $O(\sqrt[M-1]{(N)})$ sub-units at the next lower level. For LAD, M=4 (LAD-Panel-Module-Detector), and N=2016, which would suggest a branching factor of ~13. However, additional constraints result from the number of panels (6 for the consortium design) and the rectangularity of the module (for packing reasons). The values adopted (16 detectors per module, and 21 modules per panel) are close to this scheme. This approach gives a module with the parameters shown in Table 1.

Table 1 Module parameters

| Parameter | Value (no margin applied) |
|---|---|
| **Mass** | 6.05 Kg |
| **Volume** | 0.544 x 0.333 x 0.065 $m^3$ |
| **Effective area (at 8 keV)** | 0.079 $m^2$ |
| **Power dissipation** | 7.5 W |

This design gives a module which is a convenient size and mass for the activities described at the top of this section.

Design drivers for the module include the following:

**Alignment:** An MPO pore has a triangular response function in angle space. This means that the response of a single pore detector would be highly sensitive to pointing drifts and jitter. When an assembly of pores is considered, pore-to-pore misalignments must be taken into account. For LAD this involves the following: pore-to-pore within a tile, tile-to-tile within a module, module-to-module within a panel, and panel-to-panel. The overall LAD angular response is a superposition of all the (variously aligned) pore response functions. Thus, alignment effects are critical in both achieving the required on-axis collecting area, and in the pointing requirements. For this reason, a detailed analysis of alignment and pointing effects was carried out in the assessment phase, and budgets and requirements established for alignment effects and spacecraft (s/c) pointing. Note that alignment of SDDs is not important, only the collimators. Thermal gradients also factor into the alignment budgets.

**Filter:** The SDDs, being silicon detectors, in addition to the wanted X-ray sensitivity, are also sensitive to visible and UV. A metallised polymer filter [5] is placed in front of the detectors, which transmits X-rays, but rejects UV, VIS and IR (IR rejection for thermal reasons). The module has to provide a mount for the filter, allowing for vibration and acoustic loads during launch.

**SDD Temperature:** SDD dark current is a significant noise source, especially at End Of Life (EOL), when the detectors will have acquired radiation damage. The module has to provide an efficient cooling coupling from the SDDs to the radiator on the back surface.

**Radiation:** The SDDs are also sensitive to minimum-ionising radiation, both as spurious counts, and as radiation damage. The module provides shielding to the SDDs, via the MPOs and collimator frame on the front side, and the detector frame, electronics, radiator and a 200 micron layer of lead on the back side.

## 3. MODULE MECHANICAL/THERMAL DESIGN

As mentioned in the previous section, the basic LAD detection element is the Detector, composed of SDD+FEE+Collimator+Filter. Figure 3 shows the 4x4 SDDs mounted in the detector frame, while the collimators are mounted in the collimator frame which therefore has the critical alignment surfaces. Each of these detectors has a single SDD tile mounted on top of the front end electronics (FEE) board with wire bonded electrical connections running around the edges. Flexi circuit ribbons run from one end of the FEE board to Hyperstac connectors on the back end electronics (BEE) and HV boards. More details on the construction of the Module are provided in [1].

The collimator over each individual detector is a single micro-pore plate (MPP) 110.0 x 71.5 x 5 mm that fits into a recess in the aluminium alloy collimator frame. Due to the difficulty of achieving adequate collimator frame stiffness in the weight and space available, a new brazed frame design has been investigated. This frame features hollow tubular box section members to give maximum stiffness for minimum weight and can be fabricated from machined channel section halves that are then dip brazed together and finish machined.

The plan is then to place the whole collimator assembly face down onto a purpose made surface plate, following a concept that has been conceived to try to prevent any “locked in torque” in the final assembly that could induce distortion of the module and compromise the co-alignment between the different MP collimating tiles.

The detectors are mounted in a monolithic aluminium alloy frame that supports each detector unit on four points with sufficient compliance to cater for CTE variation between the frame and the detector FEE PCB. For each detector there is a conductive link via the structure to the radiator. The SDD detector assembly itself consists of the SDD, FEE PCB and the flying leads that connect with either one of the MBEE boards and HV units. This assembly is oriented such that the integration of each module is kept as modular as possible, without creating a need to introduce different interfaces between MBEE and the detector assembly related to their relative position inside the module. For this reason the MBEE is split into two PCBs. Each of the two sub-modules consists of a group of 8 detectors with all related front end electronics. The harnesses are then combined into a single interface on the outside of the module box.

Two Back-End Electronics (BEE) PCB assemblies with HV supplies attached are then mounted behind the bank of detectors, each BEE/HV assembly covering eight detectors. An aluminium alloy radiator panel of 2 mm thick screwed directly to the back of the detector module body then completes the build of the lower detector module assembly. The radiator thickness is adopted to minimise the thermal gradient across this radiator and therefore optimises the detector temperature over each orbit.

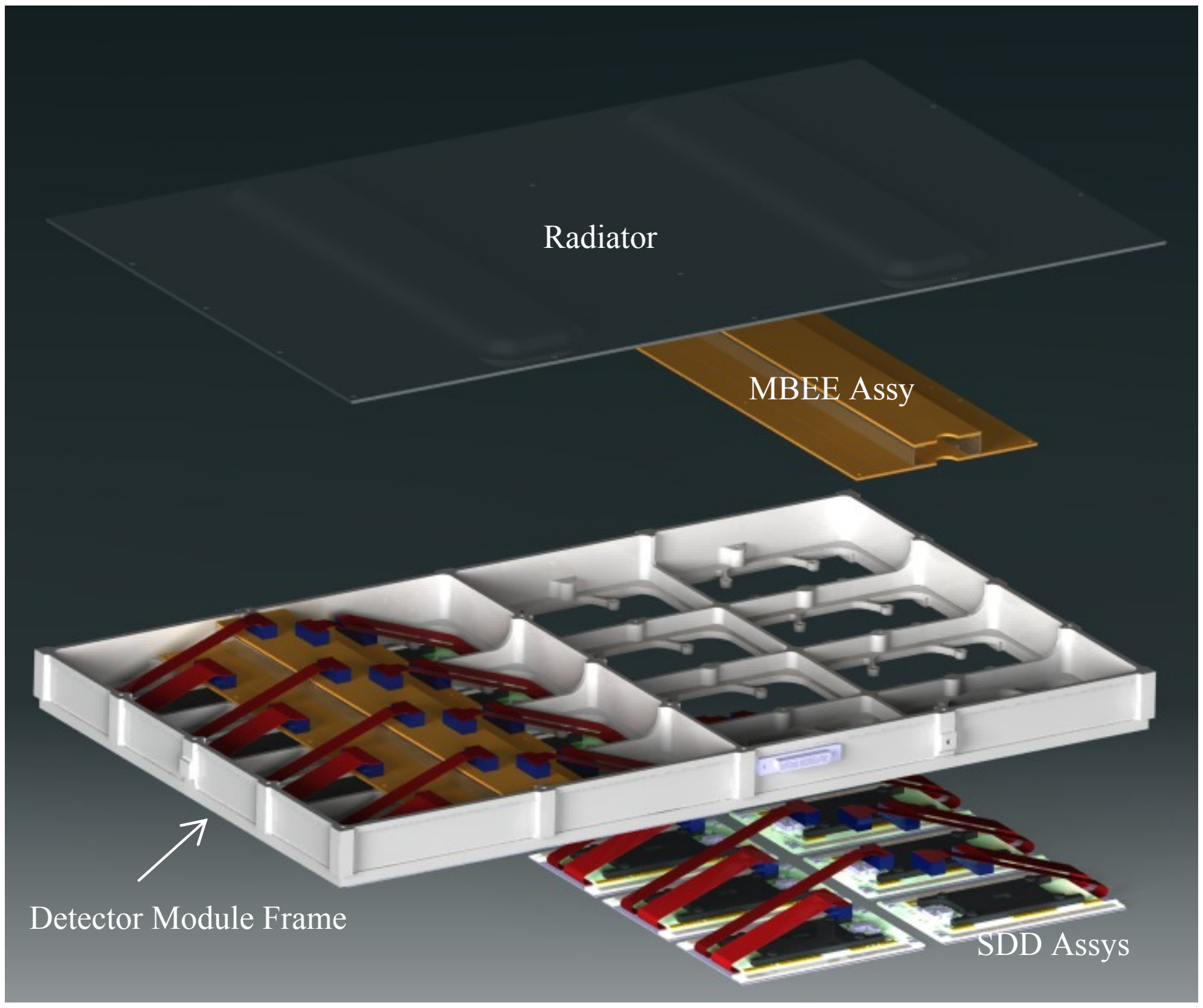

Figure 3 Exploded view of the lower detector module

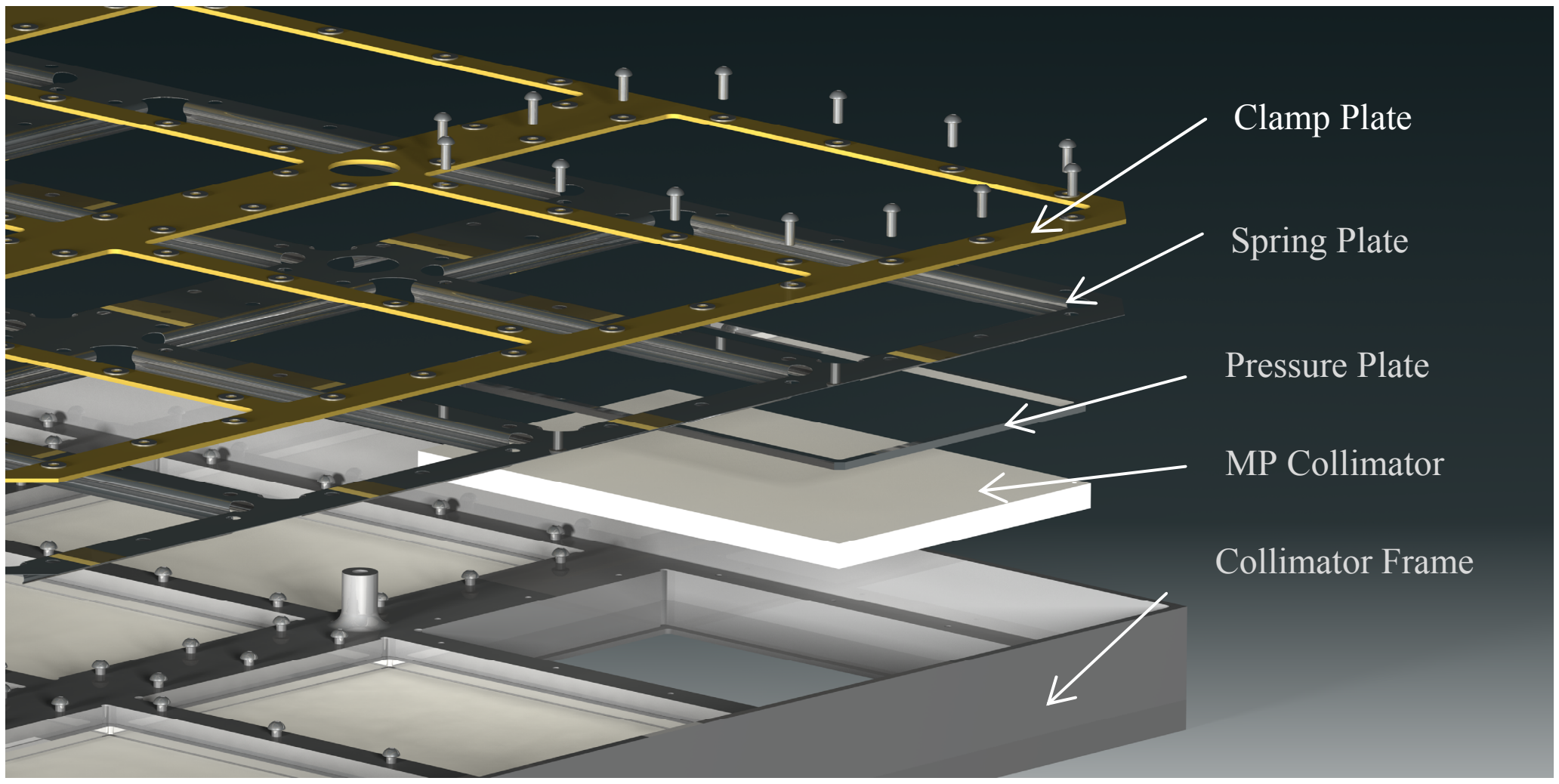

Figure 4 Exploded view on back of collimator

The overall mechanical design of the FEE assembly (SDD+PCB+Carbon Fibre Reinforced Plastic back support) has also been studied in detail during the assessment phase. Particular attention has been paid to: i) the mechanical alignment of the ASIC pads, ii) thermo-mechanical stresses that can arise due to the combination of very low survival temperatures (down to -70°C, on orbit) and the stacking of different materials in the FEE assembly, and iii) mechanical stresses due to the launcher.

## 4. THERMAL TRADE-OFF

Dark current is a significant noise source for LAD, so detector temperature is a key parameter. The spacecraft orbit is near equatorial with an inclination of 2.5 degree and an altitude of 600 km. The main sources of heat are therefore the Sun when the spacecraft is not in eclipse, the Earth as a black body and the albedo scattering of sunlight. In addition to this the detectors, ASICs predominantly, and the front end electronics dissipate about 7.5 W of heat as well.

Temperature gradients across the panel also need to be modeled, in order to assess resulting misalignments which impact the on-axis effective area.

The thermal model of the LAD module has also evolved over the past two years, addressing changes in orbit and thermal optical base line. Currently the module consists of a box holding the detectors and electronics, with a frame in front of it holding the MPOs, and a radiator at the back. The MPO tiles make up 70% of the surface of the frame. The thermo-optical properties of the tiles used for the study were based on the work done by Leicester University and ESA for the MIXS instrument on Bepi Colombo. The solar absorption is defined as 0.62 and the infra-red emission as 0.91. Our analysis uses these values as we expect the LAD MPOs will have values very near these measured by ESA for the MIXS tiles.

The box is effectively covered in second surface mirror tape, where possible. This means that the radiator is covered, as well as the frame holding the MPO tiles. This was introduced to optimise heat rejection combined with minimising solar heat absorption. Furthermore, the thermal link between each detector and the radiator has been maximised. An extensive design trade was performed to assess optimal configurations for the module from a thermal point of view: a range of steady state analyses have been performed using a detailed thermal model, aimed at sizing gradients across the detectors as well as the structure, MPO and radiator.

For the trade-off the following was included.

a) Thermal optical properties for the radiator

b) Thermal optical properties for the frame holding the MPO tiles

c) Optical filter location

For different pointings, the radiator or the MPO frame can be facing the sun, so it is important to absorb the minimum heat possible while at the same time trying to radiate away as much heat as possible. Therefore it is obvious that a material with low solar absorption (alpha) combined with a high infra-red emission (epsilon) is a good choice. So we are looking for low alpha/epsilon ratios. There is little we can do about the thermo-optical properties of the tiles themselves, they have an alpha/epsilon ratio of 0.68. Visually, the tiles look white due to the high lead content of the glass, and this helps to lower the alpha/epsilon ratio.

The radiator and the exposed surface of the MPO frame need a thermal optical control surface. Commercially available, and space approved, white paints have alpha/epsilon ratios of 0.3 and below, down to 0.1. The white AZ-93 paint in use on the space station is conductive and achieves an alpha/epsilon ratio of 0.17. Commercially available space approved second surface mirror tape also can achieve 0.17.

Apart from the micro-pore optics there is also a need to filter out the visible light into the ultra violet. To reject visible and UV, a thin layer of aluminium needs to be present between the incoming light and the detector itself. Modeling shows that a thickness of 80 nm is effective, but this is too thin to be unsupported over a length of 120 mm. There are two ways this thin layer can be supported, either directly by the MPO tile or deposited on a separate thin layer of Kapton. The Kapton backing of the aluminium needs to be very thin, a maximum thickness of 1 micron is allowed, otherwise it will start to absorb too much of the incoming low energy X-rays.

In the thermal model it is assumed that the total power dissipation inside the module is 7.5 W. We used a reduced thermal model with only a few thermal nodes to perform a parameter study. Orbital averages gave for the nominal thermal model (representing the baseline LAD design with the micro-pore optics facing the sun at a 45° angle) the heatloads as shown in the pie diagram below.

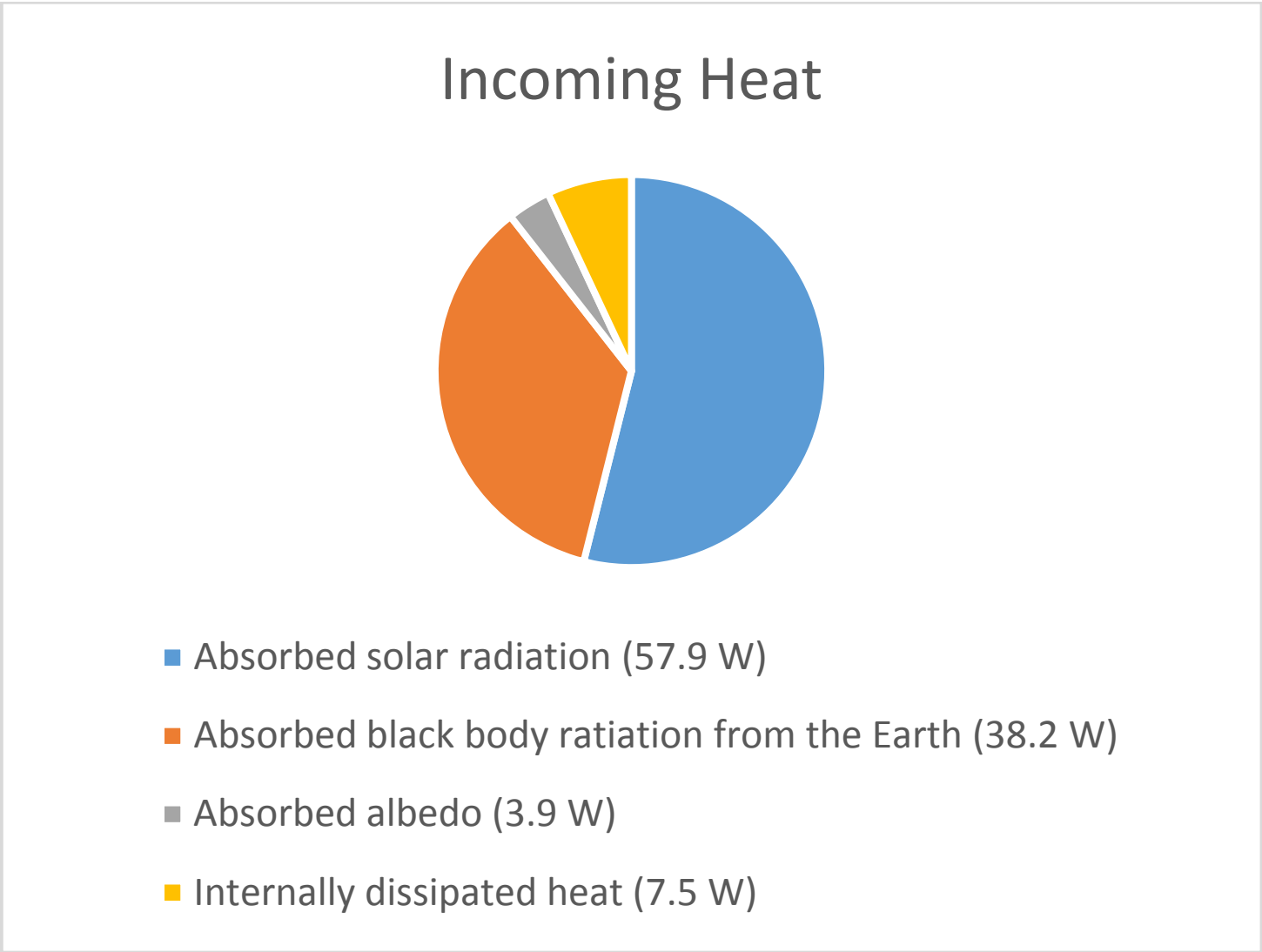

Figure 5 Distributed heat inside each LAD module for worst orbital orientation.

Therefore each LAD module needs to radiate away 107.4 W for this orbit at the worst case tilting angle for the whole array of modules. In fact the angle is +15° beyond what is required. This makes for a total of 13.5 kW for the whole array at this angle. The field of regard for the LAD is between -70° and +30°. At 0° the MPO line of sight is perpendicular to the Sun. Figure 6 shows the two extreme angular positions of the field of regard. The LAD needs to be able to operate between these angles. The current thermal model shows that LAD maintains the required detector temperature with some margin for these angles (so that in principle it could operate even at slightly larger angles).

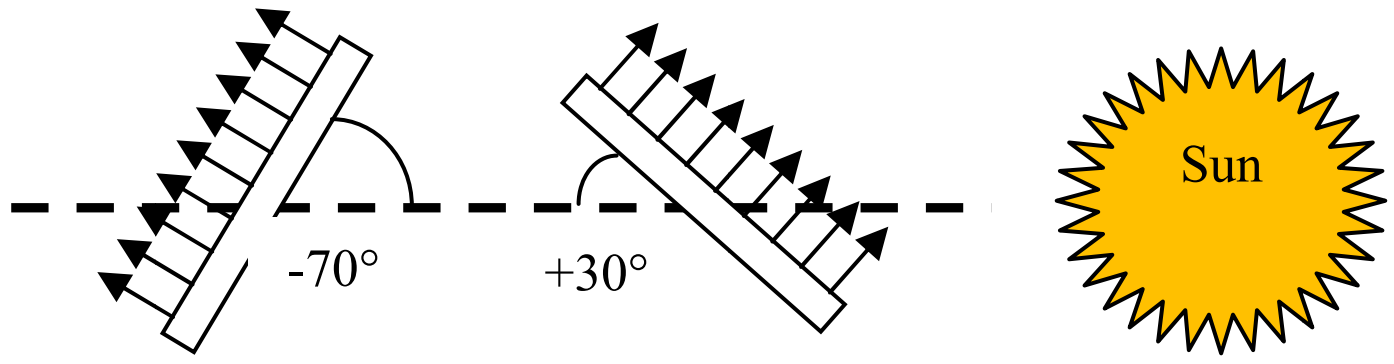

Figure 6 Required extreme pointing angles for LAD

The nominal design for the LAD is to have SSM tape bonded to the outside of the box (including the radiator), and the optical filter is nominally a separate item from the MPO tile. This provides a slightly better thermal performance than having the filter as part of the MPO tile, although the benefit is small. Using white paint can be slightly better from a thermal point of view than using SSM tape: although the alpha/epsilon ratio is very similar, the paint has the advantage that it emits infrared heat better and therefore cools down more in the eclipse part of the orbit. The other advantage of paint is that it is lighter. However, the SSM is baseline at the moment since it has more heritage with the potential prime contractors and protects better against atomic oxygen.

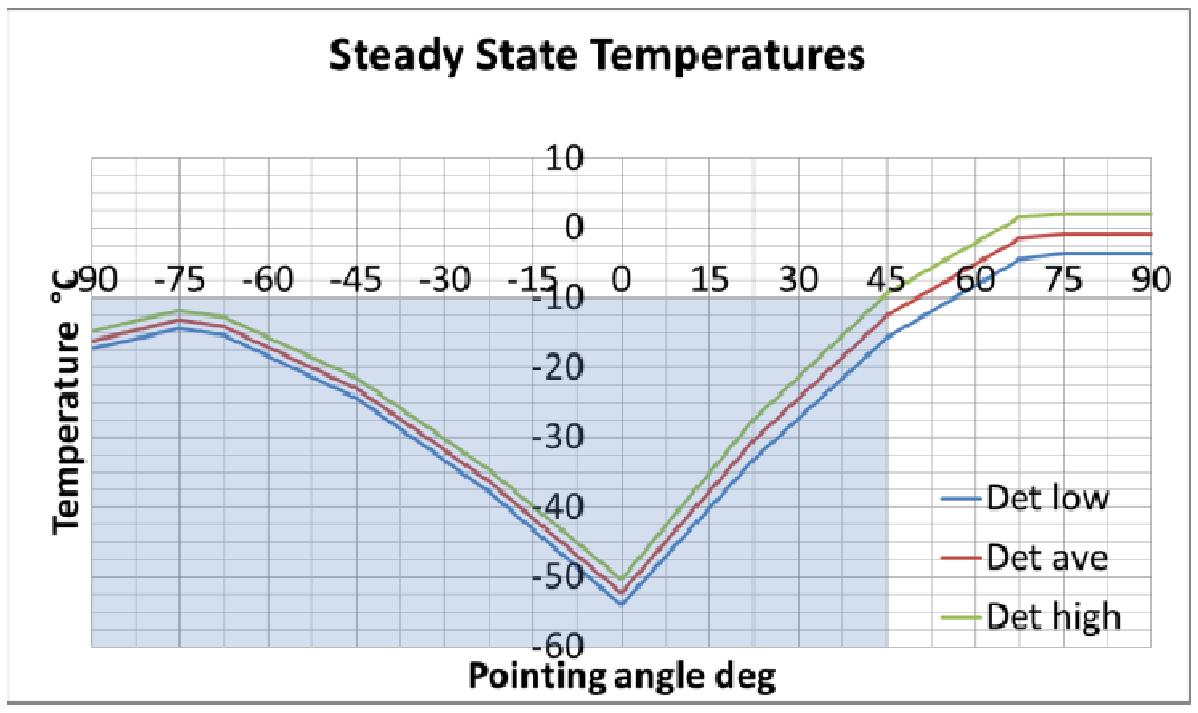

Figure 7 Detector temperature as function of pointing for the nominal configuration

Variations in the radiator surface and the filter location from the baseline design do not give a hugely different thermal response with significant shifts in detector temperature. White paint on the radiator instead of second surface mirror tape lowers the temperature of the detectors by as much as 3.5°C. Variations in filter locations and coating of the filter make a difference, but by less than 1°C on average.
So in conclusion, the LAD has been optimised with detectors operating between -10°C and -50°C, the lower end of this operating range is close to the extreme lower end of MIL-spec standard parts for electronics.

Design alternatives, such as paint instead of SSM tape or the location of the optical filter are interchangeable and can be traded in later if required to save mass or cost or allow for a different supplier to mitigate risk.

In order to increase margin to help cope with additional heat radiated from the spacecraft (which has not been included in this trade) there is a need for additional shading or cooling. This could be achieved by using a sun shade for the array of modules which could be deployed as part of the panel deployment. In addition to this, the radiator at the back of each module can be extended somewhat in size. There is room in the overall design to do this but it will cost extra mass.

Another concern is the thermal gradient across the detectors. There is no room in the design of the module at present to optimise this other than keeping the structure as uniform as possible from a thermal point of view. This has been achieved by adding a spine through the centre of each module. This spine has a similar thermal through-conduction as the sidewalls and adds stiffness to the module as well. The spine splits each module neatly into two sets of 8 detectors.

Hereafter the results of the detailed thermal model are shown, illustrating the thermal gradients we can expect.

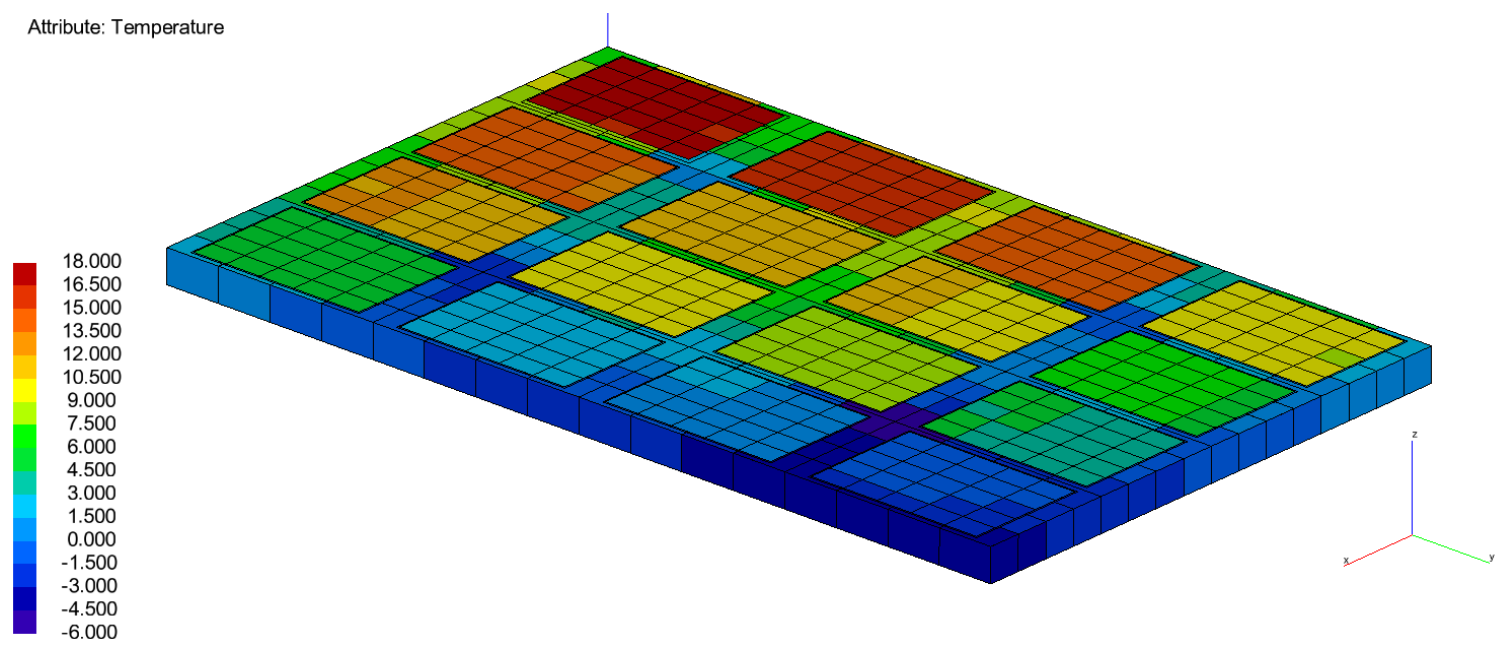

Figure 8 MPO frame and tiles with thermal gradient, 45° tilt towards the sun

As can be seen in Figure 8 above, there is a significant gradient across the MPO tiles at the +45° tilt towards the sun. The frame has a significant gradient as well, 14°C between the extreme corners.

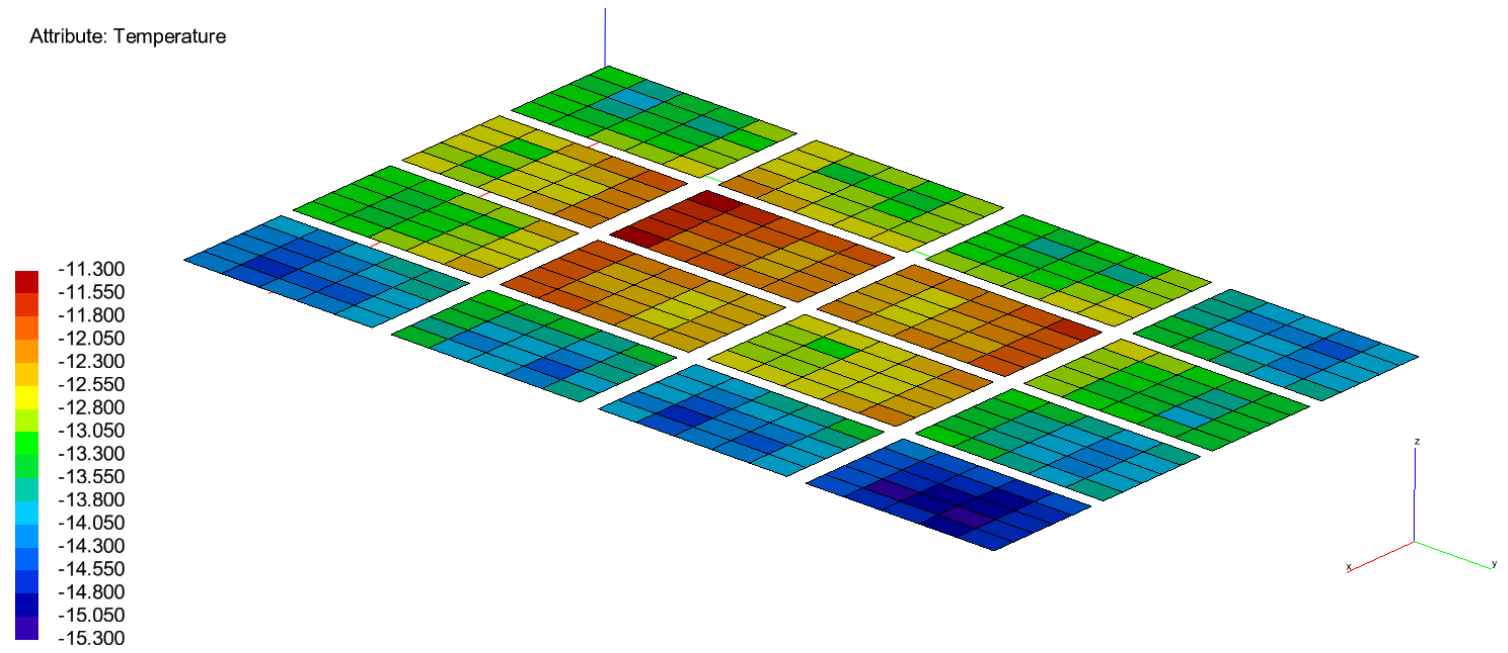

Figure 9 The detector temperatures when tilted 45° towards the sun

As can be seen, the gradient noticeable along the tiles (Figure 8) is not reflected strongly in the detector temperatures (Figure 9). There is a gradient and the influence of the gradient across the MPO tiles is noticeable, but the effect has been minimised. For this analysis each detector has been split into 25 elements. The detectors are supported and thermally linked at four nodes, each near the four corners of each detector. The uniformity of the temperature of each detector itself is very good.

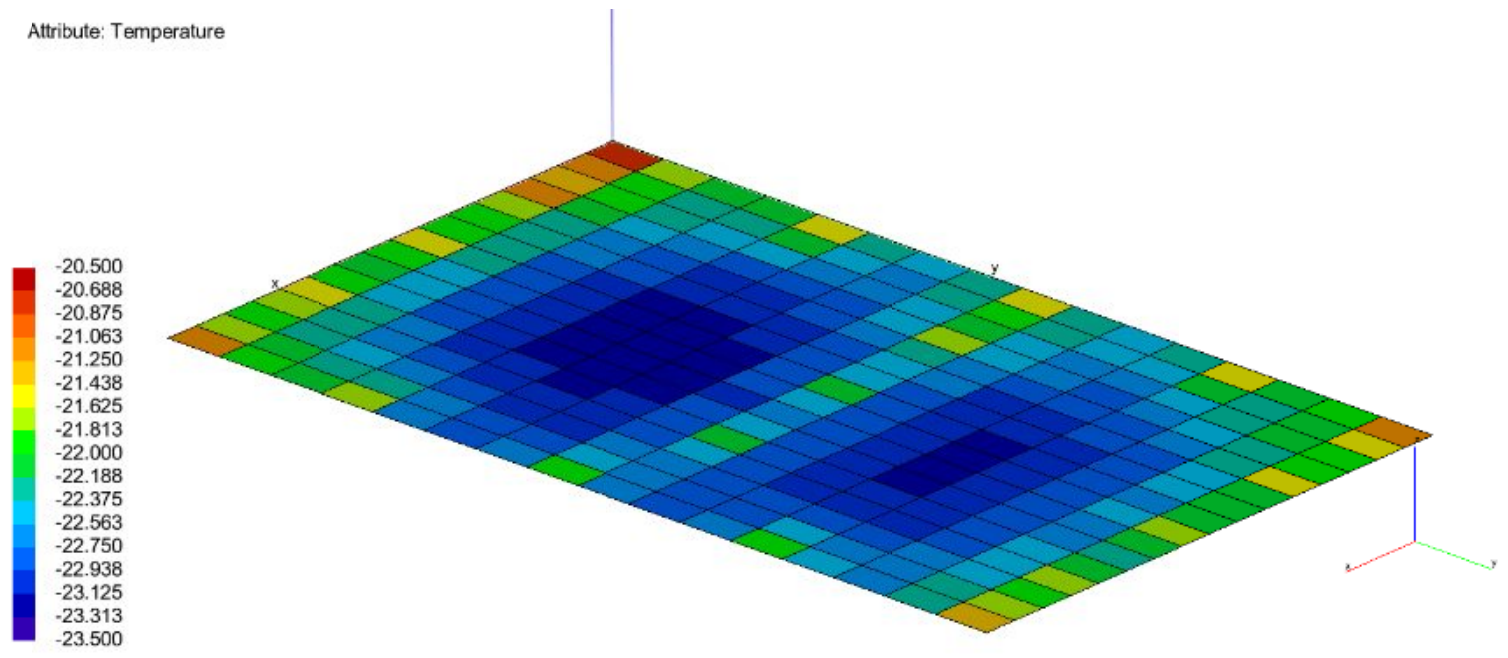

Figure 10 Radiator gradient

Figure 10 shows the gradient across the radiator at the extreme pointing angle of 45° away from the Sun (i.e. radiator at 45° towards the sun). It is clear from the graph that the gradient is minimal and if the radiator would be extended in size, maintaining the same footprint of the box, there is room for more cooling without increasing the gradient. The local hot spots are where the radiator is attached to the main module structure.

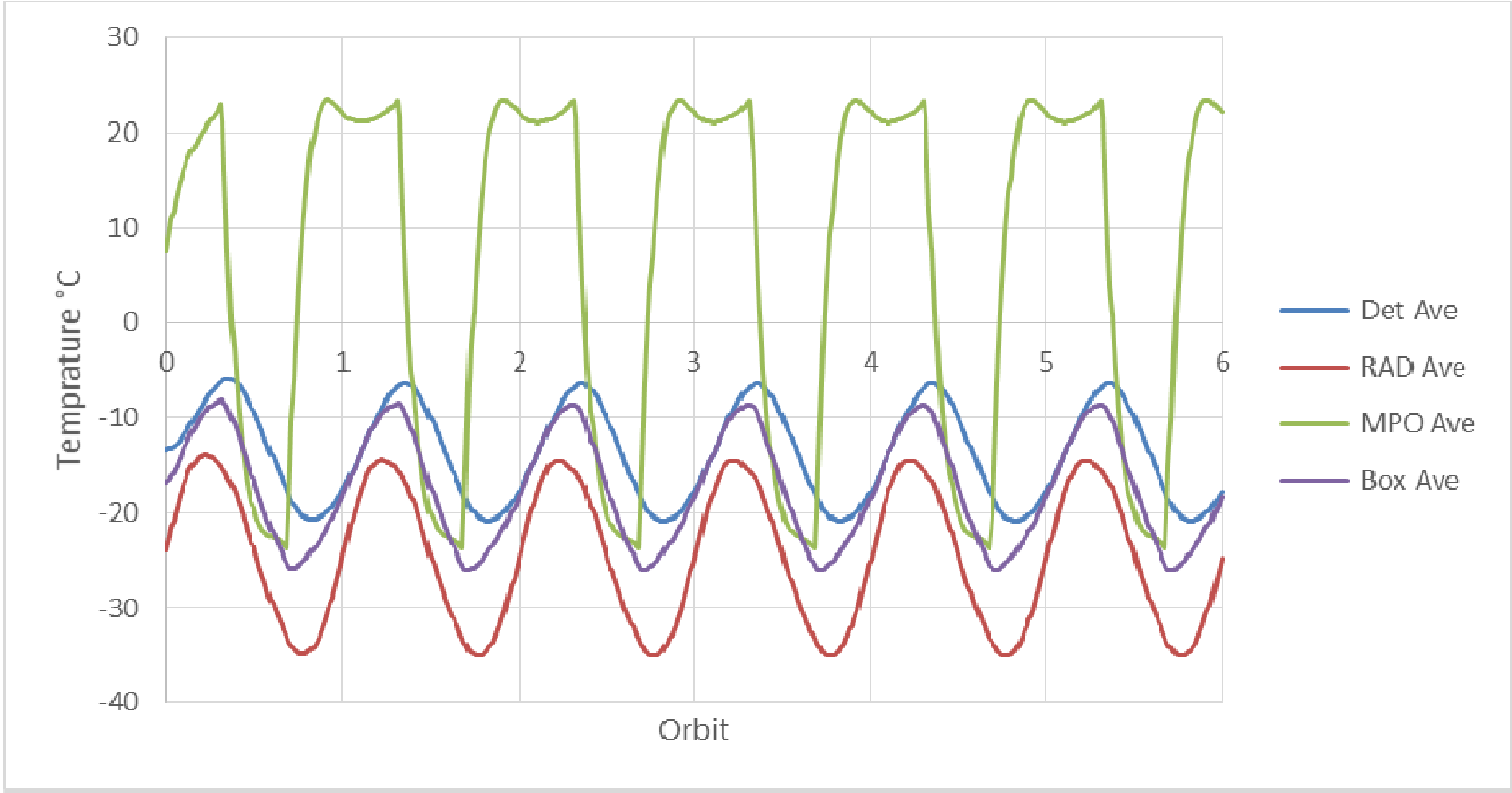

Figure 11 Transient temperatures

It should be noted that the gradient shown in Figure 9 is the average temperature over the orbit. At this extreme angle some of the detectors will drift in temperature above -10°C for part of the orbit. Figure 11 shows the results of a reduced thermal model of variation over several orbits.

The orbital transient shown in Figure 11 starts at a steady state point (orbital average) and then continues for 6 orbits. After the first orbit the model has already almost stabilised. The MPO tiles which are pointing towards the sun at 45° are fluctuating significantly in temperature as is to be expected. Since the main structure (Box) of the module is used to cool down the detectors (Det), the structure temperature is slightly below that of the detectors. The radiator (RAD) is the lowest in temperature at all times as expected. For the 30° sunpointing angle the detectors are all the time below -10°C.

### 4.1 Thermo-elastic trade-offs

Thermo-elastic deformation within each module has to be considered, taking into account material incompatibilities as well as general deformation due to thermal gradients. Due to temperature changes, parts with materials with dissimilar Coefficients of Thermal Expansion (CTE) that are connected will deform and generate thermo-elastic stresses.

To minimise this effect, materials are matched as far as possible with regard to the CTE, and materials are selected that help to minimise thermal gradients.

There are three areas of concern, the CTE mismatch between the MPO and the frame, the mismatch between the SDD and the PCB (detector sub assembly) and the thermal gradient over the thickness of the module. We will discuss here the MPO-frame mismatch and the thermal gradient over the module.

The MPO tile has a CTE of 8.8 micro/K, which is close to that of titanium, titanium 6AL4V has an average CTE of 8.1 micro/K between -40°C and +40°C. Therefore in the operating temperature range this titanium alloy is a good match. Aluminium in the same temperature range has an average CTE of 22.3 micro/K (6082T6), which is 2.5 times higher. So the obvious choice, from a thermo-elastic point of view, for holding the MPO tiles in place is this titanium alloy. There is however a price to pay when using titanium, which is mass. For a titanium frame to work, it should be mounted on top of a box that has a similar CTE, otherwise the frame will be stressed due to temperature changes. The frame itself is large and needs at least 4 locations where it is bolted down on the box. Hence a titanium frame would require a titanium box to support it.

Titanium is 70% denser than aluminium, and the specific stiffness of aluminium (the ratio between density and stiffness) is close to that of titanium, so in theory one should be able to design a frame with the same mass either in titanium or in aluminium. However the detailed design of the frame and the box does not allow for that efficiency. The cross-sections in aluminium are already thin and on the edge of what is acceptable from a radiation shielding point of view. It would be very hard to come close, in mass, using titanium.

But for LAD, there is a more important difference between aluminium and titanium: aluminium conducts heat a factor of 30 better than titanium. Conduction of heat away from the detectors is critical, so this drives the decision to use aluminium for the main structure and the MPO frame, and work around the significant difference in CTE between the MPO tiles and the frame. This CTE mismatch raises the need for clearances between the tiles and the frame itself. The required clearance is small, 0.15 mm over the length of the MPO, but significant for the survival of the tile.

Since we cannot use a rigid connection between the frame and the tiles, the tiles will be held in place by clamps. The clamps are in practice custom-made leaf-springs, which provide for enough preload to prevent the tiles from moving out of plane during vibration. The springs do not clamp directly on the brittle tiles, load spreaders are used to introduce the clamp force as gently as is practical. The interface between the tiles and the bare frame again comprises a load spreader to prevent local surface defects (machining marks) from causing stress concentrations. Due to the large number of tiles required we try to avoid interface surface polishing, the machining will be as clean as possible but we will ultimately rely on a load spreader to absorb local and more global surface imperfections.

The in-plane location of the tiles does not rely on the preload provided by the clamps: friction clamping would require an undesirable additional preload on the brittle tiles. Instead, viton tubes will run along the edge of the tiles, providing for sufficient preload to maintain the location of the tiles but at the same time leaving enough flexibility for the tiles to ‘breathe’ when cooling down without cracking them.

The thermal gradient over the height of the module itself is minimised as mentioned before by using aluminium. Aluminium alloy has a very low CTE/thermal conduction ratio. In theory materials like SiC or AlBeMet will perform better, but the use of a ceramic composite for the frame and main structure would be a wrong choice from a manufacturing point of view. The intricate details of the box and the frame are not suited to manufacture using a ceramic material. Beryllium, as part of the AlBeMet alloy is best avoided at all times due to excessive cost and machining difficulties when taking into consideration health and safety.

### 4.2 Collimation and Alignment

As described previously, the critical alignment issue for LAD is to maintain the coalignment of the individual collimator pores.

The purpose of the collimator is to provide LAD with its angular sensitivity to celestial point X-ray sources at energies within its operating band and inside its field-of-view, while effectively suppressing all source X-rays incident at larger angles. The collimator design (see [2] for details) is based on the technology of microchannel plates (MCPs) and thus drawing on the heritage of the EXOSAT (1983-6) MEDA and GSPC detectors and on the much more recent developments for the BepiColombo Mercury Imaging X-ray Spectrometer (MIXS).

The science requirements which have driven the LAD collimator design are (i) the field-of-view (ii) the required transparency of the collimator at high (ie 30-50 keV) X-ray energies and (iii) a possible requirement for a “flat-top” addition to the basic triangular collimator response function, describing X-ray transmission versus off-axis angle. The basic collimator units will consist of 8 x 11 $cm^2$ tiles, fabricated from lead glass, with parallel 83micron square cross-section channels, 5mm long. The specification for the pore-to-pore co-alignment is 1 arcmin. The open area fraction will be 70%, giving a low collimator mass of ~5$kg.m^{-2}$.

The overall angular response function of the LAD is a convolution of the triangular response of an individual pore with the pseudo-Gaussian distribution of misalignment effects. These misalignment effects occur at various levels:

- Pore-to-pore within a tile: A result of MPO manufacturing tolerances. Care is taken in the tile mounting to avoid torsionally stressing the tile, which could introduce systematic pointing misalignments, or at worst cracking of the brittle MPO tiles.
- Tile-to-tile within a module: Determined by the manufacturing tolerances of the collimator tray. Care is taken to minimize thermal gradients across the module which could produce pointing misalignments. The thermal modelling described in the previous section has been used to analyse this effect, as well as ensuring that the detector temperature is within an acceptable range.
- Similar arguments apply to the alignment of modules within a panel. In terms of the module design, care has to be taken that the module provides well-toleranced mounting points for attachment to the panel. Fiducial marks will also be provided on the module to allow optical checking of alignment during integration into the panel.
- Panel-to-panel alignment is equally important, but is outside the scope of this paper.

An alignment-related trade-off is that the misalignments reduce on-axis effective area, but flattens the top of the triangular response function. This flattening is beneficial in that pointing drifts/jitter then have a reduced effect in terms of effective area variation, an important consideration for studies where frequency analysis is important. Table 2 shows the predicted values for the various misalignment effects, and Figure 12 shows the effect of the convolution on the effective area v. angle.

Table 2 Alignment assessment

| **Consortium responsible** | **Error (arcmin, RMS)** |
|---|---|
| a) MPO internal | 1.0 |
| b) MPO-to-module | 0.43 |
| **RSS Subtotal** | 1.09 |

| **Industry responsible** | |
|---|---|
| c) Module-to-panel | 0.59 |
| d) Thermo-elastic panel (10 C) | 0.39 |
| e) Panel-to-optical-bench | 0.34 |
| f) Panel 1-g | 1.3 |
| g) Optical Bench | 0.5 |
| RSS Subtotal | **3.12** |
| | |
| **RSS TOTAL** | **4.55** |

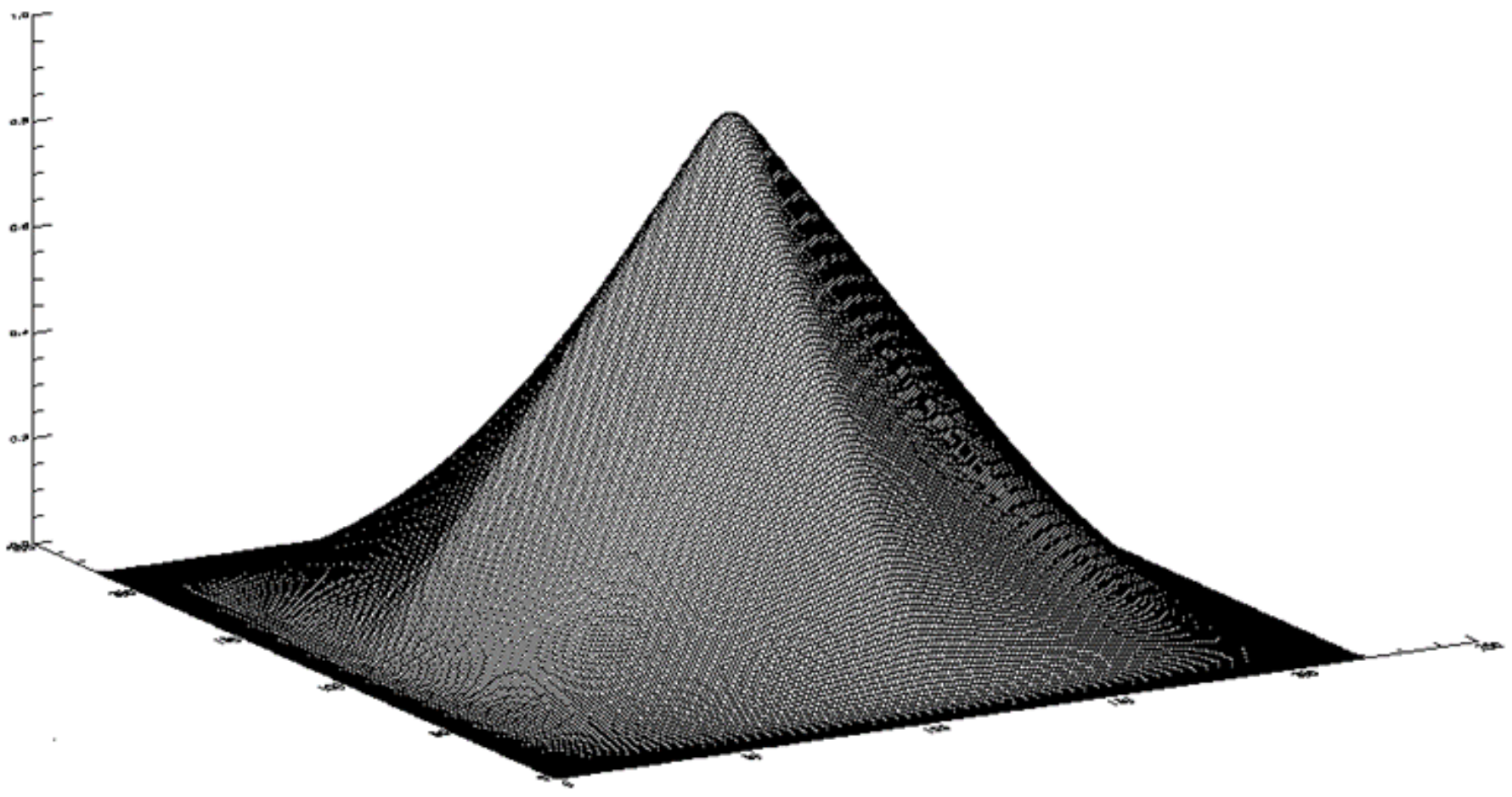

Figure 12 Effective area v. angle: Pore triangular response modified by pseudo-Gaussian misalignment effects

### 4.3 Filter

At present the baseline is to have an optical filter mounted between the detector and the micro pore plate. Due to envelope constraints this filter will be mounted on the collimator frame, above each detector. Design of these frames needs to be progressed paying attention to purging the micro pore tiles as well as the acoustic loading of these particularly thin filters (a test assembly has been produced for early acoustic testing and the brazed frame structures are well advanced and have been successfully used for vibration testing and proof of analytical results).

## 5. MODULE ELECTRICAL DESIGN

### 5.1 Overview

Figure 13 shows the electrical architecture from instrument level down to module level, and Figure 14 shows the electrical architecture within a Module (more detail is available in [2] and [6]). At the top level are the spacecraft Data Handling Unit (DHU) and Power Distribution Unit (PDU, supplying 50 V DC). These are connected to the Instrument Control Unit (ICU, one operational and another in cold redundancy). The ICU is connected to 12 Panel Back-End Electronics units (PBEE, two per panel in this configuration). Each PBEE is connected to 10 or 11 Modules (the module electrical unit is the Module Back-End Electronics, MBEE). Each MBEE is connected to 16 sets of Front-End Electronics (FEE), where the FEE is the electronics unit for a single detector.

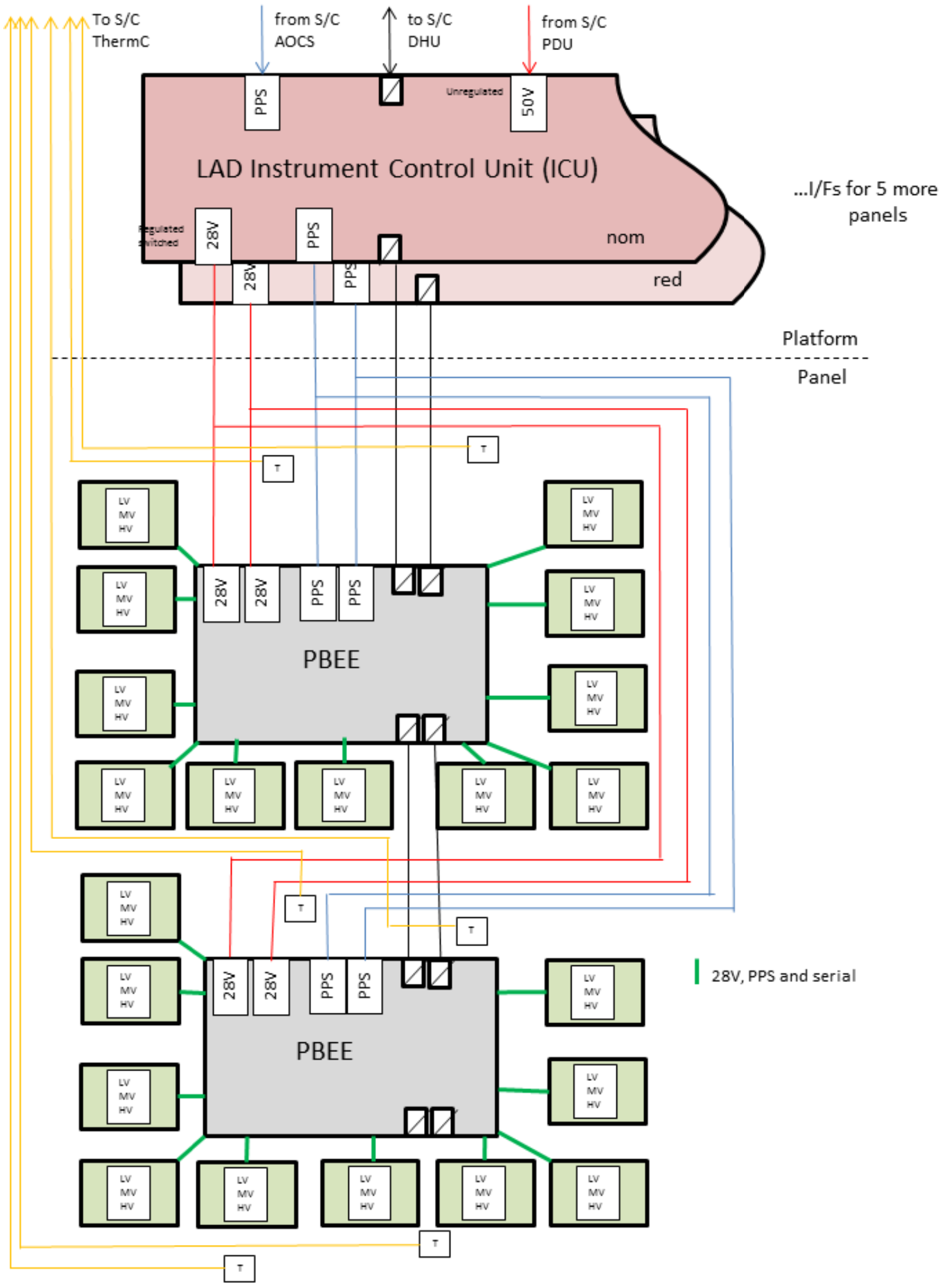

Figure 13 Electrical configuration of the LOFT-LAD, module and upstream.

There are two aspects to the electrical architecture, power supply and signals, as described below.

**Power supply and distribution:** As shown in Figure 13 and Figure 14, a 50 V DC supply (or 28 V, TBD) is provided by the spacecraft. This is converted down to 28 V in the ICU, and the 28 V is distributed to each PBEE. Each PBEE can be switched off in the event of power irregularities. The PBEE distributes the 28 V to each module, and on the module power supply board, the SDD HV, SDD MV, and detector digital and analogue LV supplies are generated.

**Signals:** The bulk of the data traffic is the flow of digitized events from the detector ASICs through the MBEEs, through the PBEEs and to the ICU. Commands (e.g. instrument modes) and data values (e.g. digital offsets) pass in the opposite direction. Each level of data link has been considered in terms of the protocol to be used.

The individual ASICs for each detector (14 per detector, digitizing a total of 224 anodes per detector) send event data to the MBEE on a simple custom serial link within the module. Data rate per ASIC is low (8.5 cts/s for a 1 Crab source), and the communications are highly deterministic, so a simple interface is adequate, without the protocol overheads of, say, SpaceWire.

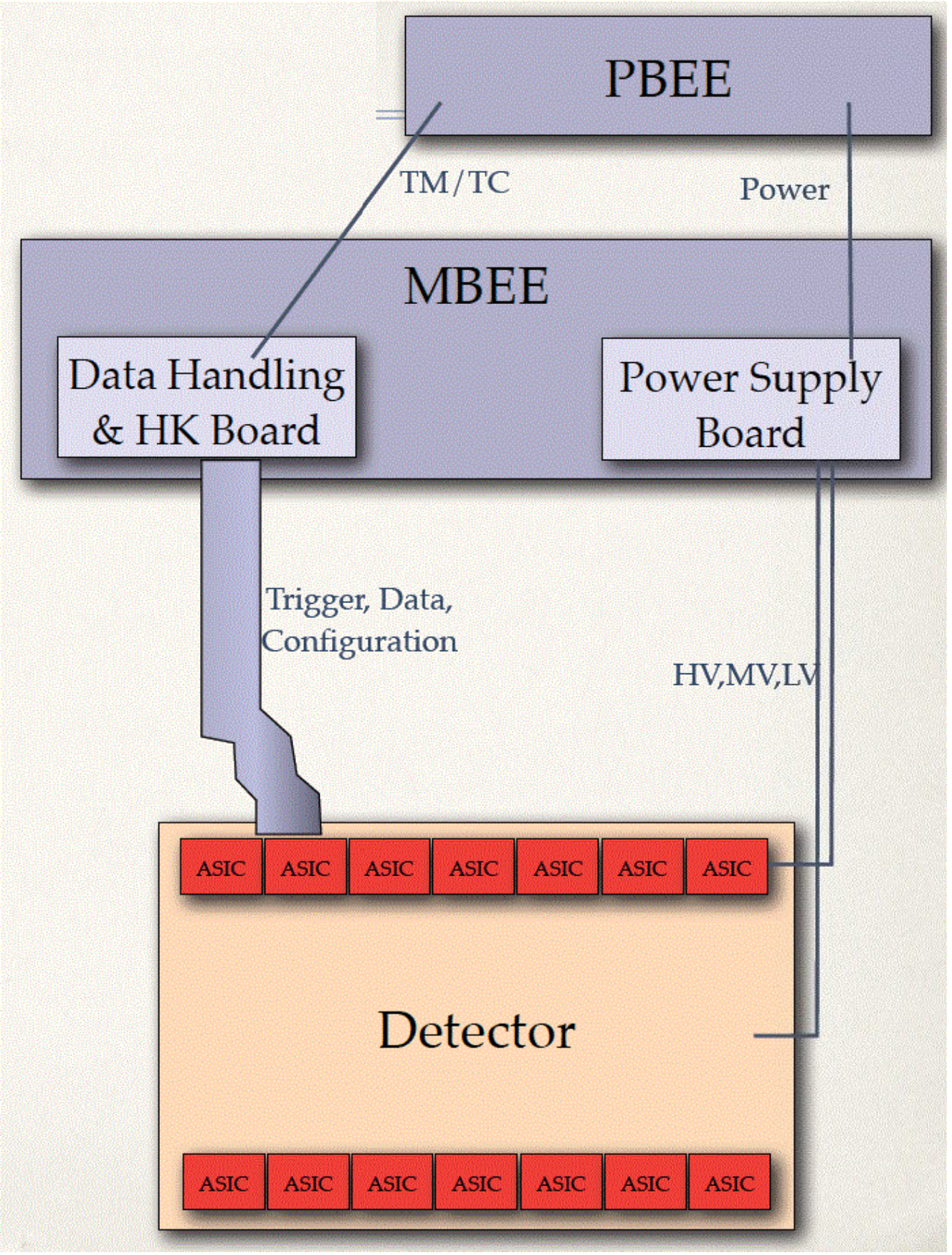

Figure 14 Electrical configuration of the LOFT-LAD, module and downstream. For clarity, only one PBEE-module link and one MBEE-detector link are shown. In the consortium design, each PBEE is connected to ~11 modules (MBEE), and each MBEE is connected to 16 detectors.

The MBEE takes the data from all the ASICs and passes it on to the PBEE (after dealing with e.g. multi-anode events, and time-tagging the data). The data rate is still low (1920 ct/s for 1 Crab) so again a custom serial interface is used, this time using LVDS signals, due to the longer harness from Module to PBEE.

Each PBEE receives the events from ~11 to 21 Modules (depending on the PBEE architecture chosen), performs some compression and reformatting, and passes the data to the ICU. At this level, the data rate is higher (~20000 to 40000 ct/s for 1 Crab), with each event providing 24 bits of data (amplitude, timing and flags), and ICU-PBEE communications are more complex, so SpaceWire is chosen for this interface.

## 5.2 Trade-offs/optimisation

During the assessment phase, various electrical-related trade-offs were considered before the scheme described above was adopted, as described below.

**PBEE location:** In order to reduce power dissipation on the panel, in principle the PBEE could be mounted on the s/c bus. However, this increases the harness length from the PBEE to Modules (power, data and clock), and in particular the amount of harness running across the panel-to-bus hinge, which has to be deployed after launch. Simulations of panel temperature showed that the temperatures are acceptable with the PBEE on the panel, so it was decided to place the PBEE on the panel in order to save harness mass and hinge mechanical resistance.

**Power conversion location:** It is desirable to minimize power dissipation in the module, so that the detector temperature can be kept low, to keep dark current down. For this reason, the option of performing the power conversion in the PBEE was considered. Another benefit is reduced duplication of conversion circuitry in the modules, saving module mass. One downside of this approach is that LV then has to be distributed to the modules, and distributing power at LV requires larger cross-section conductors, negating the mass saving mentioned above. Another downside is the increased risk of breakdown and electrical noise when distributing HV across the panel. These factors led to the adoption of performing all conversion from 28 V within the module. Other benefits of this approach are that power dissipation across the panel is more uniform, reducing misalignment effects due to thermal gradients, and that the modules are more stand-alone, improving the testability and replacement philosophy.

# 6. CONCLUSIONS

The LOFT LAD module design has evolved throughout the LOFT Assessment Phase in response to various constraints and boundary conditions ranging from universal concerns (e.g. mass, power) to highly instrument-specific issues (e.g. collimator tile-to-tile alignment and designing to ease module-to-module alignment). This paper has described some of the salient trade-offs and optimisations which have resulted in a robust module design to take forward to future LOFT developments.

# 7. ACKNOWLEDGEMENTS

Because the Module is so inter-disciplinary, thanks are due to a huge number of people across the consortium for their efforts in improving the Module design, including the detector experts, electronics and mechanical designers and technicians (including Simon Hemsley and the MSSL mechanical workshop staff for their efforts in prototyping), project managers, AIV and calibration experts and more. The authors would like to express their thanks to all.

The work of the UCL-MSSL and Leicester groups is supported by the UK Space Agency. The Italian team is grateful for support by ASI (under contract I/021/12/0-186/12), INAF and INFN. The work of SRON is funded by the Dutch national science foundation (NWO). The work of the group at the University of Geneva is supported by the Swiss Space Office. The work of IAAT on LOFT is supported by Germany’s national research center for aeronautics and space DLR. The work of the IRAP group is supported by the French Space Agency.